# Staggered Comb Reference Signal Design for Integrated Communication and Sensing

Rui Zhang
Department of Electrical Engineering
University at Buffalo
Buffalo, NY, USA
rzhang45@buffalo.edu

Shawn Tsai
Advanced Comm. Technology
Comm. System Design
MediaTek USA Inc.
San Diego, CA, USA
shawn.tsai@mediatek.com

Tzu-Han Chou
Advanced Comm. Technology
Comm. System Design
MediaTek USA Inc.
San Diego, CA, USA
tzu-han.chou@mediatek.com

Jiaying Ren
Advanced Comm. Technology
Comm. System Design
MediaTek USA Inc.
San Diego, CA, USA
jiaying.ren@mediatek.com

*Abstract*—Ambiguity performance is a critical criterion in radar sensor design, which indicates the ambiguities arising from multiple targets estimation and detection. We considered a requirement-driven selection of OFDM reference signal (RS) patterns based on ambiguity performances for bi-static sensing in integrated communication and sensing with minimal modifications of current RSs. An RS pattern with a staggering offset of a linear slope that is relatively prime to the RS comb size is suggested for standard-resolution sensing algorithms to obtain the best ambiguity performances. Moreover, an extended guard interval design is proposed to increase the maximum time delay, that is inter-symbol interference (ISI) free using post-FFT sensing algorithms. The proposed techniques are promising to extend the distance and speed without ambiguities and ISI for sensing.

*Keywords—Integrated communication and sensing, reference signal, 6G, bi-static sensing, delay and sum, Periodogram Algorithm*

## I. Introduction

Recently, integrated communication and sensing (ICAS) has been envisioned to support diverse applications potentially in the fifth generation advanced (5G-A) and the sixth generation (6G) era [1], such as activity or object detection, recognition, and tracking. The integration mutually benefits communication and sensing by saving system resources. One research direction is communication centric ICAS, which realizes sensing functionality in a system primarily designed for communications. In such a system, it is preferred for sensing function to reuse current wireless communication system's waveforms, e.g., orthogonal frequency division multiplexing (OFDM) adopted by 5G New Radio (NR). Detecting Doppler shift and time delay by using OFDM waveform has been extensively studied for radar sensing [2-6].

For bi-static sensing with unknown data payload, RS is straightforward for sensing receivers to estimate time delays and Doppler shifts of targets [5,7,8], proportional to their range and the radial velocity, respectively. Studies to leverage current LTE and 5G RS including positioning reference signal (PRS) for radar sensing have been given in [5,7,8]. Most communication RSs were designed for channel state information and communication receiver demodulation, and 5G PRS has a special design of staggering offset sequence for delay estimation but not for both delay and Doppler estimation. In radar sensing, resolutions and ambiguities of both delay and Doppler are two important criteria for system design. High-resolution sensing algorithms such as compressive sensing and MUSIC can also improve sensing resolution but show higher computational complexity than standard-resolution algorithms. Fundamentally the total bandwidth and the time duration of a sensing signal determines the achievable delay and Doppler resolutions, respectively. We will show in next section that, tuning RS resource elements (REs) density can adjust the resolutions but at the same time change ambiguities in both delay and Doppler domains. In this paper, based on standard-resolution sensing algorithms, we will analyze the ambiguity performance of comb RS patterns with different RE densities. We will show how the staggering offset design through multiple symbols could help to improve its ambiguity performance. The standard resolution algorithms considered here include delay-and-sum [9] and periodogram algorithms such as two-dimensional fast Fourier transform (2D FFT) algorithms [10]. We will first investigate RS patterns similar to current 5G NR for sensing. A new design is also proposed for improved ambiguity performance with minor changes to RS patterns in the current 5G communication system. Moreover, we introduce an extended guard interval design for comb RS to enable ISI removal beyond cyclic prefix (CP).

## II. General Comb RS Patterns with Uniform Symbol Spacing

There are a few types of 5G NR RS: demodulation RS (DMRS), channel state information RS (CSI-RS), tracking RS (TRS), phase tracking RS (PTRS), and positioning RS (PRS) [11].

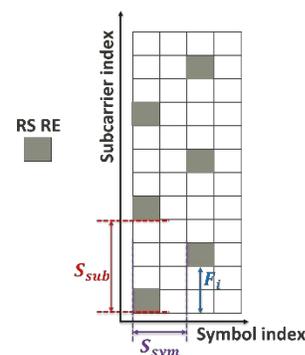

Figure 1. Illustration of general comb RS pattern

These RS patterns can be characterized by Fig. 1 with the following parameters: (1) $S_{sub}$ (unit in subcarrier numbers) is the RS frequency domain spacing of REs and $S_{sub} \geq 2$ for comb RS. $S_{sub}$ is also called comb size, which determines the frequency domain RE density. (2) $S_{sym}$ (unit in symbol num-bers) is the time domain separation of the RS symbols (RE density in the time domain), and (3) $F_i$ (unit in subcarrier numbers) is the frequency domain RE offset (or the staggering frequency offset) of the $i$-th RS symbol. In this paper, based on such a comb RS structure with uniform RE spacing and symbol spacing from 5G NR, we analyze impacts of $S_{sub}, S_{sym}$ as well as staggering frequency offsets $\{F_i\}$ on the sensing ambiguity performance.

The comb RS has several benefits. First, it boosts energy per RE for better coverage. Second, it maximizes the delay resolution with smaller number of REs spanning across the entire channel bandwidth, compared to subband RE-block allocation. Third, it enables either different base-station PRS multiplexing or data RE insertion. However, the comb RS introduces delay ambiguity at a fractional symbol level for standard resolution algorithms. Therefore, the comb RS pattern over multiple symbols with a cycle of staggering offsets within the coherent processing interval (CPI) is adopted in 5G PRS to preserve up-to-one-effective-OFDM-symbol unambiguous delay. The approach was to pair two comb RS symbols and put their staggering offsets one-half-comb-size ($S_{sub}/2$) apart, as shown in Fig 2(b). Then, if more OFDM symbol pairs can be allocated, offsets further sweep over full $S_{sub}$ REs to cover the entire frequency range, e.g., two pairs of staggering offsets (0, 2) and (1, 3) for $S_{sub} = 4$ as shown in Fig 2(c). The design was considered for scenarios where PRS resources may not be available for more than two symbols within the CPI. Paired one-half-comb-size staggering offsets double the maximum unambiguous time delay given by a single comb RS symbol, resulting in better performance in ranging for positioning.

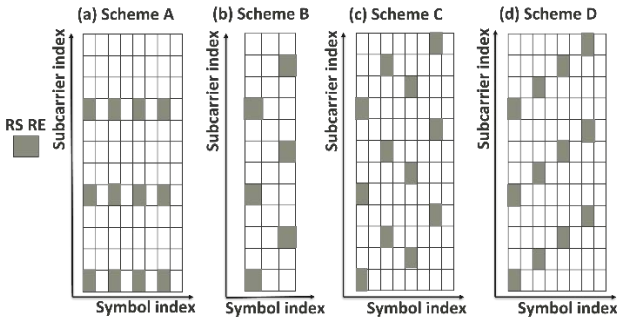

Figure 2. Instances of RS patterns

Nonetheless, sensing requires another look at other RS pattern possibilities. Specifically, with Doppler frequency shift added to the estimated variables, ambiguities arise in the 2D delay-and-Doppler plane. Under a fixed number of REs, larger $S_{sub}$ and $S_{sym}$ increase occupied bandwidth and time duration, which enhance range (time delay) and velocity (Doppler shift) resolutions, respectively. However, such resolution improvement by tuning $S_{sub}$ and $S_{sym}$ comes with costs of reduced maximum unambiguous delay and Doppler shift. In the following sections, we will analyze trade-offs between different RS patterns in terms of ambiguity performance.

Several RS patterns with uniform symbol spacing are shown in Fig. 2 and categorized by Scheme A, B, C and D for the remaining part of the paper. Fig. 2(a) illustrates an example of Scheme A with $S_{sub} = 4, S_{sym} = 2$, which is similar to 5G PDCCH DMRS or TRS. This pattern keeps the same RE locations across all RS symbols, i.e., $F_i$ stays constant for any $i$-th RS symbol. Fig. 2(b) depicts an example of Scheme B, which is similar to partial PRS (subgroup of a full staggering cycle) with $S_{sub} = 4, S_{sym} = 2, F_0 = 0$, and $F_1 = S_{sub}/2$,. In 5G NR, partial PRS has the format of staggering two RS symbols' REs by half an even comb size, (i.e., for $F_0 = f, F_1 = f + S_{sub}/2$). Fig. 2(c) shows an instance of Scheme C with $S_{sub} = 4, S_{sym} = 2$, using the 5G PRS staggering offset sequence. When $S_{sym} = 1$, the RS pattern is a 5G PRS. Fig. 2(d) illustrates an instance of the proposed Scheme D with $S_{sub} = 4$, $S_{sym} = 2$, and $p = 1$, which is linear-slope staggering with the slope relatively prime to the comb size. The staggering offset sequence is $F_i = \mod(p \cdot (i-1), S_{sub})$ for $i = 0, \cdots, S_{sub} - 1$, where $p$ is relatively prime to $S_{sub}$.

## III. AMBIGUITY FUNCTION OF DELAY AND SUM

The ambiguity function (AF) of a time-domain signal $s(t)$ based on delay and sum is expressed as [9]

$$A(\tau, f_d) = \frac{1}{E_s} \left| \int_{-\infty}^{\infty} s(t) e^{(j2\pi f_d t)} s^*(t - \tau) dt \right|, \quad (1)$$

where $E_s = \int_{-\infty}^{\infty} s(t) s^*(t) dt$, $f_d$ is the Doppler frequency shift, and $\tau$ is the time delay. In general, the design criterion is to minimize $A(\tau, f_d)$ in the areas where $(\tau, f_d) \neq (\tau^*, f_d^*)$, where $\tau^*$ and $f_d^*$ are the nominal true time delay and Doppler. To simplify the description in the following sections, $(\tau^*, f_d^*)$ is set to (0,0) without loss of generality (WLOG) as AF is a form of RS pattern's impulse response in the 2D domain. Large values of $A(\tau, f_d)$ exceeding a radar dynamic range at $(\tau, f_d) \neq (0, 0)$ are undesired side peaks. These side peaks limit the unambiguously detectable range of time delay and Doppler frequency shift, and we name it 2D maximum unambiguity hereafter. When there are side peaks, the design criteria then become how to enlarge the 2D maximum unambiguity, or how to design the waveform such that side peak locations $(\tau, f_d)$ are as far away from (0, 0) as possible.

We now derive a general expression of AF based on the general comb RS patterns in Section II. Let $T_s$ be the effective OFDM symbol duration (the reciprocal of the subcarrier spacing (SCS)), $T_{cp}$ be the CP duration, $T = T_{cp} + T_s$ be the CP-added OFDM symbol duration, and $X_i$ be the frequency-domain RE scrambling sequence of the $i$-th comb RS. The $i$-th RS symbol sample vector of length $N$ over duration $T_s$ is denoted as $\mathbf{Y} = \{Y(n)\}_{n=0}^{N-1}$, where the $n$-th sample is

$$Y(n) = \sum_{\kappa=0}^{N/S_{sub}-1} X_i(\kappa S_{sub} + F_i) e^{\frac{j2\pi(\kappa S_{sub} + F_i)n}{N}}, \quad (2)$$

Observe that

$$Y\left(n' + \frac{Nl}{S_{sub}}\right) = \sum_{\kappa=0}^{N/S_{sub}-1} X_i(\kappa S_{sub} + F_i) e^{\frac{j2\pi(\kappa S_{sub}+F_i)\left(n'+\frac{Nl}{S_{sub}}\right)}{N}}$$
$$= Y(n') e^{j\left(\frac{2\pi F_i l}{S_{sub}}\right)}, \quad (3)$$

where $n' = 0,1,2,\cdots, N/S_{sub} - 1$ and $l = 0,1,2,\cdots, S_{sub} - 1$. Thus $\mathbf{Y}$ can be further divided equally into $S_{sub}$ subsets, where the $l$-th subset is denoted by $\mathbf{Y}_l$, each of length-$(N/S_{sub})$ and

$$\mathbf{Y}_l = \mathbf{Y}_0 e^{j\left(\frac{2\pi F_i l}{S_{sub}}\right)}, \mathbf{Y}_l = \{Y(n)\}_{n=lN/S_{sub}}^{(l+1)N/S_{sub}-1}. \quad (4)$$

Let the time sequence of CP-added OFDM symbol be $\mathbf{Z} = \{Z(n)\}_{n=0}^{N'}$, and the $n$-th sample of $\mathbf{Z}$ over the CP-added duration $T$ be

$$Z(n) = \sum_{\kappa=0}^{N/S_{sub}-1} X_i(\kappa S_{sub} + F_i) e^{\frac{j2\pi(\kappa S_{sub}+F_i)(n-N_{cp})}{N}}, \quad (5)$$

where $N' = (N + N_{cp})$. As time samples of one OFDM symbol possess $S_{sub}$ subsets that are repetitive with piecewise phase rotation as shown by Eq. (4), strong side peaks of a delay-and-sum receiver happen at $\tau = l \cdot T_s/S_{sub} = l \cdot N/S_{sub} \cdot (T_s/N)$, where $l \cdot N/S_{sub}$ is the beginning sample index of the $l$-th subset of Eq. (3). Let $M$ be the number of RS symbols, the AF at time delay $l \cdot T_s/S_{sub}$ and Doppler frequency $f_D$ by delay-and-sum over $M$ RS symbols is expressed by:

$$A\left(\frac{T_s l}{S_{sub}}, f_d\right) = \left|\sum_{i=0}^{M-1} \sum_{n=iS_{sym}N'+\frac{Nl}{S_{sub}}}^{iS_{sym}N'+N'-1} Z(n) Z^*\left(n - \frac{Nl}{S_{sub}}\right) e^{\frac{j2\pi f_d nT}{N}}\right|. \quad (6)$$

Assuming $\mathbf{X}_i$ preserves constant envelope in the time domain (e.g., frequency-domain Zadoff-Chu sequence), we obtain

$$A\left(\frac{T_s l}{S_{sub}}, f_d\right)$$
$$= \left|\sum_{i=0}^{M-1} e^{-\frac{j2\pi F_i l}{S_{sub}}} \sum_{n=iS_{sym}N'}^{iS_{sym}N'+N'-1-\frac{Nl}{S_{sub}}} Z(n) Z^*(n) e^{\frac{j2\pi f_d\left(n+\frac{Nl}{S_{sub}}\right)T}{N}}\right|$$
$$= \left|S \cdot e^{\frac{j2\pi f_D lT}{S_{sub}}} \sum_{i=0}^{M-1} e^{-\frac{j2\pi F_i l}{S_{sub}}} \sum_{n=iS_{sym}N'}^{iS_{sym}N'+N'-1-\frac{Nl}{S_{sub}}} e^{\frac{j2\pi f_d nT}{N}}\right|, \quad (7)$$

where $S = Z(n)Z^*(n)$ is a constant. Delay-and-sum algorithm requires the time sequences derived from $\mathbf{X}_i$ to have desirable autocorrelation properties. Doppler frequency is detected by a variable-$f_d$-phase-rotated version of the time sequence. In the case of $f_d = 0$ (e.g., positioning), the staggering offset sequence $\{F_i\}$ directly affects the side peak locations by

$$A\left(\frac{T_s l}{S_{sub}}, 0\right) = \left|S \cdot \left(N' - \frac{Nl}{S_{sub}}\right) \sum_{i=0}^{M-1} e^{-\frac{j2\pi F_i l}{S_{sub}}}\right|, \quad (8)$$

For instance, $M = S_{sub}$, $\{F_i\}_{i=0}^{S_{sub}-1} = \{0,1,\cdots, S_{sub}-1\}$, the side peaks along the axis of $f_d = 0$ in the time delay range $(0, T_s)$ can be eliminated as $\sum_{i=0}^{M-1} e^{-\frac{j2\pi F_i l}{S_{sub}}} = 0$ for all $l \in \{1,\cdots, S_{sub}-1\}$. AF characteristics derived from delay-and-sum algorithm over various RS patterns will be examined next in this section.

*A. Scheme A: Patterns similar to 5G PDCCH DMRS or TRS*

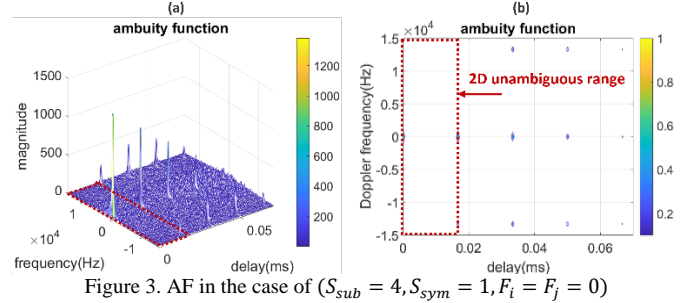

Figure 3. AF in the case of $(S_{sub} = 4, S_{sym} = 1, F_i = F_j = 0)$

As mentioned in in Section II and based on Eq. (7) and Eq. (8), Scheme A ($F_i \equiv$ constant without staggering) has 2D AF side peak locations at $(\tau = lT_s/S_{sub}, f_d = k/S_{sym}T)$, where $l, k \in \mathbb{Z}$ (set of integers), $0 \leq l \leq S_{sub}, -S_{sym} \leq k < S_{sym}$ and $(l, k) \neq (0,0)$. Note that there is an exception for delay-and-sum to be free of ambiguities at $(\tau = 0, f_d = \pm 1/T)$ because there is no discrete-time Fourier transform to create frequency domain aliasing. If the RS symbols do not have time interruption (i.e., $S_{sym} = 1$), the maximum unambiguous Doppler frequency for delay-and-sum is only limited by sampling rate $N'/T$ because there is no frequency-domain convolution with a non-impulse. Those side peaks creating ambiguities for delay and sum results are shown by the example in Fig. 3 with $S_{sub} = 4$, $S_{sym} = 1$, $F_i \equiv 0$, under SCS=15 kHz. The contour plot of Fig. 3(a), as given in Fig. 3(b), will be used for ambiguity performance analysis onwards. The 2D maximum unambiguity, the red dotted rectangle in Fig. 3(b), designates the 2D AF maxima without any ambiguities in both time delay and Doppler frequency shift domain. As shown in Fig. 3(b), undesired side peaks fall outside the red rectangle, meaning no ambiguities exist in the designated 2D range.

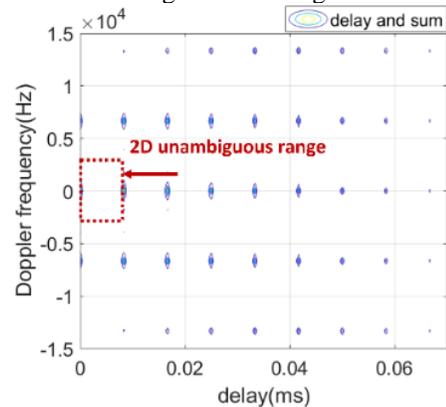

Figure 4. AF of comb RS with $(S_{sub} = 8, S_{sym} = 2, F_i \equiv 0)$

Fig. 4 shows another example with $S_{sub} = 8$, $S_{sym} = 2$, and $F_i \equiv 0$. Note that the increased comb size ($S_{sub}$ from 4 to

8) reduces the maximum unambiguous time delay, and the interruption of RS continuation in time ($S_{sym} = 2$ instead of 1) causes convolution with a gating function's frequency-domain response, creating ambiguities along the true delay's Doppler frequency axis before reaching $N'/T$.

Note that the designated 2D maximum unambiguity is not unique. Depending on application scenarios, the designation around true delay and Doppler could have different options:
1) In the case of $S_{sym} = 1$, time delay is from 0 to $T_s/S_{sub}$, Doppler frequency is from $-N'/(2T)$ to $N'/(2T)$ as shown in Fig. 3.
2) In the case of $S_{sym} > 1$, time delay is from 0 to $T_s/S_{sub}$, and Doppler frequency is from $-1/(2S_{sym}T)$ to $1/(2S_{sym}T)$ as shown in Fig. 4.

However, the maximum unambiguous delay in this case is always restricted to $T_s/S_{sub}$.

### B. Scheme B: Patterns similar to partial PRS

In the case of scheme B, half-comb-size staggering is incorporated, creating alternating $S_{sub}$-subset ambiguities. That is, for even integers $l_1$, side peaks occur at $(\tau = l_1 T_s/S_{sub}, f_d = k_1/S_{sym}T)$, for $0 \leq l_1 \leq S_{sub}$, $-S_{sym} \leq k_1 \leq S_{sym}$, $k_1 \in \mathbb{Z}$, and $(l_1, k_1) \neq (0,0)$, where forementioned $(\tau = 0, f_d = \pm 1/T)$ exception still holds. For odd integers $l_2$, side peaks occur at locations $(\tau = l_2 T_s/S_{sub}, f_d = (1/2 + k_2)/(S_{sym}T))$, for $0 \leq l_2 \leq S_{sub}, -S_{sym} \leq k_2 \leq S_{sym}, k_2 \in \mathbb{Z}$, and $(l_2, k_2) \neq (0,0)$. Fig. 5 shows the half-comb-size-staggering example of $F_1 = 0$, and $F_2 = 2$ under $S_{sub} = 4$ and $S_{sym} = 1$,. Such RS pattern extends the maximum unambiguous delay from $T_s/S_{sub}$ to $2 \cdot (T_s/S_{sub})$ compared with scheme A. Depending on application scenarios, the designated 2-D maximum ambiguity, beside the vertically long dotted rectangle in Fig. 5 as analyzed before, can be opted for a an extended maximum unambiguous delay from 0 to $2T_s/S_{sub}$, and a Doppler range now from $-1/(4S_{sym}T)$ to $1/(4S_{sym}T)$. Such a new option is shown by the horizontally wider dotted red rectangle of Fig. 5.

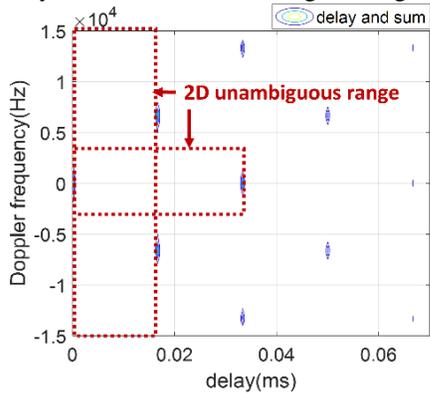

Figure 5. AF of Partial PRS with ($S_{sub} = 4, S_{sym} = 1, F_0 = 0, F_1 = 2$)

### C. Scheme C: Patterns with stagerring offset of 5G PRS

Fig. 6 shows an AF of Scheme C with $S_{sub} = 4, S_{sym} = 1$ and $\{F_0, F_1, F_2, F_3\} = \{0, 2, 1, 3\}$, which is a staggering offset sequence such that the REs sweep through the whole signal bandwidth. This PRS pattern can extend the maximum unambiguous time delay to full-symbol length $T_s$, albeit at the cost of smaller Doppler unambiguity. Based on Eq. (8), the horizontally wider dotted red rectangle of Fig. 6 shows an option to designate full-symbol delay unambiguity if the application scenario precludes large Doppler, for which $f_d$ is assumed to be between $-1/(2S_{sym}S_{sub}T)$ and $1/(2S_{sym}S_{sub}T)$.

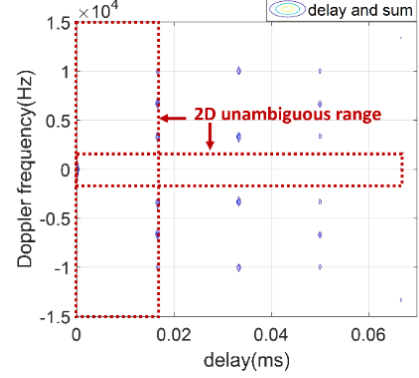

Figure 6. AF of PRS-like pattern with ($S_{sub} = 4, S_{sym} = 1$)

### D. Scheme D: Linear-slope-relatively-prime-to-comb-size staggering

Now we turn to the proposed Scheme D, where the staggering offset sequence $\{F_i\}_{i=0}^{S_{sub}-1}$ has a linear slope $p$ that is relatively prime to the comb size $S_{sub}$. Corresponding side peaks occur at locations $(\tau = l \cdot T_s/S_{sub}, f_d = p \cdot l/S_{sub}S_{sym}T + k/S_{sym}T)$ for $0 \leq l \leq S_{sub}$, $|k| \leq S_{sym}$, $(l,k) \neq (0,0)$ and the exception at $(0, \pm 1/T)$. Compared with those of aforementioned RS patterns, the proposed scheme shows more flexibility in determining 2D maximum ambiguity. The AF of $S_{sub} = 4$, $S_{sym} = 1$, and $p = 1$ is shown in Fig. 7(a) and (b) with different choices of designating unambiguous regions. Fig. 7(c) and (d) present the AF of $S_{sub} = 4, S_{sym} = 1$, and $p = 3$ (a different slope that is relatively prime to $S_{sub}$).

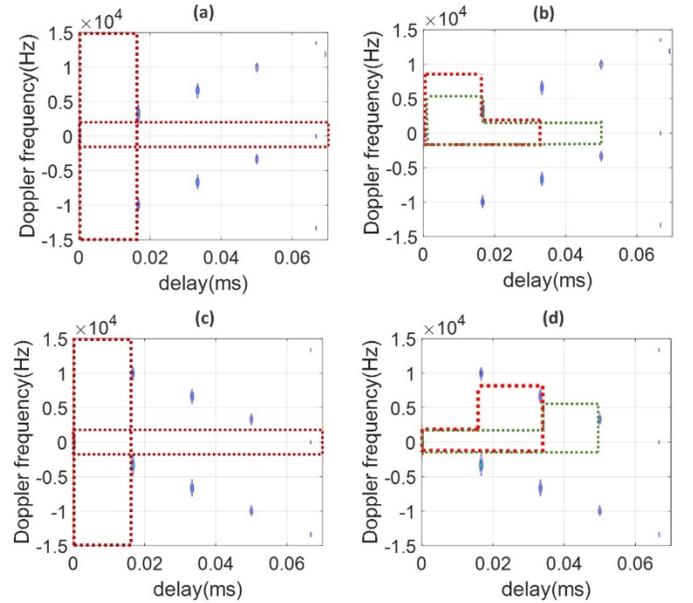

Figure 7. AF in the case of ($S_{sub} = 4, S_{sym} = 1$), proposed staggering scheme, where in (a) and (b) $p = 1$, and in (c) and (d) $p = 3$.

The choices are (taking the example with $p = 1$):
1) Comb-sized fractional-symbol delay unambiguity:
   a. For $S_{sym} = 1$, time delay is from 0 to $T/S_{sub}$, and Doppler from $-N'/(2T)$ to $N'/(2T)$, as shown in Fig. 7(a).
   b. For $S_{sym} > 1$, time delay is from 0 to $T_s/S_{sub}$, and Doppler from $-1/(2S_{sym}T)$ to $1/(2S_{sym}T)$.
2) Full-symbol unambiguous delay extended to $T_s$, with Doppler range from $-1/(2S_{sym}S_{sub}T)$ to $1/(2S_{sym}S_{sub}T)$, as shown by the horizontally wider dotted red rectangle in Fig .7(a).
3) Designated unambiguous delay from 0 to $lT_s/S_{sub}$, where $l = 2,\cdots, S_{sub} - 1$, as shown in Fig. 7(b): One designated 2D maximum unambiguity is expressed as:
$$\begin{cases} -\frac{1}{2S_{sym}S_{sub}T} < f_d < \frac{2S_{sub}-2l+1}{2S_{sub}S_{sym}T}, & 0 < \tau < \frac{T_s}{S_{sub}} \\ -\frac{1}{2S_{sym}S_{sub}T} < f_d < \frac{1}{2S_{sub}S_{sym}T}, & \frac{T_s}{S_{sub}} \le \tau < \frac{lT_s}{S_{sub}} \end{cases}.$$
Fig. 7(b) shows two examples – the red area denotes the case of $l = 2$ and the green area denotes the case of $l = 3$.
Another choice could be:
$$\begin{cases} -\frac{2S_{sub}-2l+1}{2S_{sub}S_{sym}T} < f_d < \frac{1}{2S_{sub}S_{sym}T}, & \frac{(l-1)T_s}{S_{sub}} \le \tau < \frac{lT_s}{S_{sub}} \\ -\frac{1}{2S_{sub}S_{sym}T} < f_d < \frac{1}{2S_{sub}S_{sym}T}, & 0 < \tau < \frac{(l-1)T_s}{S_{sub}} \end{cases}.$$
In summary, the staggering offset sequence $F_i = \mod(p \cdot (i - 1), S_{sub})$, where $p$ is relatively prime to $S_{sub}$, is suggested for delay-and-sum sensing algorithms as it provides more flexible 2D maximum unambiguity choices for time delay and Doppler frequency shift detection.

## IV. EXTENDED GUARD INTERVAL FOR COMB RS

When applying OFDM to ICAS, post-FFT frequency domain sensing algorithms, e.g., 2D FFT, high-resolution algorithms [2, 4,10], are conventionally limited by CP duration, which determines the maximum time delay (hence distance) without ISI. In this section, we will show that the frequency comb RS pattern with zero-power REs in between could extend the guard interval against ISI beyond CP. Flexibility of extending ISI-free time delay up to CP plus $(S_{sub} - 1)/S_{sub}$ of effective OFDM symbol length can thus be achieved by post-FFT algorithms.

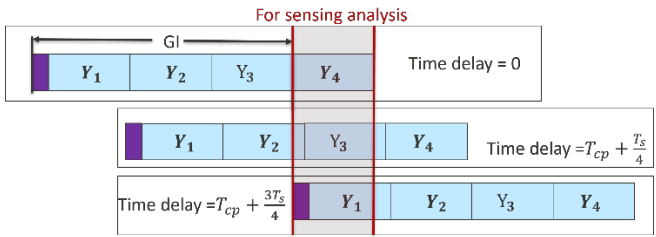

Figure 8. Sensing analysis based on $(1/S_{sub})$-symbol duration.

Assuming the Doppler shift is sufficiently small compared to the OFDM base frequency, i.e., $|f_d| < f_{max} \ll 1/T_s$ (where $f_{max}$ is typically 1/10 of SCS), we can approximate it as a constant phase rotation block-by-block over each OFDM symbol. Such an assumption was not required in the case of delay-and-sum over a continuous period for estimating the Doppler frequency variable $f_d$. However, for post-FFT sensing algorithms, it is required as time observations of Doppler frequency has been discretized at the granularity of an OFDM symbol. Recall CP-added OFDM time samples as defined in Eq. (5), an extended guard interval against sensing ISI (GI in Fig. 8) is now defined as the first $(N_{cp} + lN/S_{sub})$ samples, where $l \in \{0,1,\cdots, S_{sub} - 1\}$ ). Such extension can reuse the same CP-OFDM symbol (originally designed for shorter multipath delay communication) to longer distance sensing without resorting to a new waveform.

After removing the extended guard interval against ISI, remaining time samples of the $i$-th OFDM symbol, delayed by $N_\tau$ samples and phase-rotated by a Doppler frequency $f_d^*$, are expressed by the time sample vector of the last $(S_{sub} - l)$ subsets $G_l = \{G_l(m)\}$, for $m = 0,1,\cdots, N(S_{sub} - l)/S_{sub} - 1$. The $m$-th element of $G_l$ is

$$G_l(m) = Z\left(\frac{Nl}{S_{sub}} + N_{cp} + m - N_\tau\right) \cdot e^{j2\pi f_d^* iTS_{sym}}$$
$$= \sum_{\kappa=0}^{\frac{N}{S_{sub}}} X_i(\kappa S_{sub} + F_i) e^{\frac{j2\pi(\kappa S_{sub}+F_i)\left(\frac{Nl}{S_{sub}}+m-N_\tau\right)}{N}} e^{j2\pi f_d^* iTS_{sym}}, \quad (9)$$

where $0 \le N_\tau \le \min(N_{cp} + Nl/S_{sub}, N)$. To recover the non-zero REs from $G_l$, the receiver side should first perform the following phase de-rotation procedure before FFT, that is

$$G'(m) = G_l(m) e^{-\frac{j2\pi F_i\left(\frac{Nl}{S_{sub}}+m\right)}{N}} \quad (10)$$

for $m = 0,1,\cdots, N(S_{sub} - l)/S_{sub} - 1$. Then, after FFT of $G'$, we obtain sequence $B$ with a length of $N \cdot (S_{sub} - l)/S_{sub}$ and

$$B(\kappa') = \sum_{m=0}^{\frac{(S_{sub}-l)N}{S_{sub}}-1} G'(m) e^{j\frac{-2\pi\kappa' S_{sub}m}{(S_{sub}-l)N}}. \quad (11)$$

In Eq. (11), for $\kappa' \in \{(S_{sub} - l)w, w = 0,1,\cdots, N/S_{sub}\}$, we can write

$$B(\kappa') = (S_{sub} - l) \cdot X_i\left(\frac{\kappa' S_{sub}}{S_{sub} - l} + F_i\right)$$
$$\cdot e^{j2\pi f_d^* iTS_{sym}} e^{-\frac{j2\pi N_\tau\left(\frac{\kappa'}{(S_{sub}-l)}S_{sub}+F_i\right)}{N}}. \quad (12)$$

As for $\kappa' \notin (S_{sub} - l)w$, then $B(\kappa') = 0$. The frequency-domain sequence of the $i$-th RS symbol for sensing algorithm analysis (with a length of $N - lN/S_{sub}$) can be re-assembled to sequence $X'_i$ with the original length of $N$. For $B(\kappa') \ne 0$, we then map $X'_i(\kappa' S_{sub}/(S_{sub} - 1) + F_i) = B(\kappa')$. For RE locations other than $\kappa' S_{sub}/(S_{sub} - 1) + F_i$, $X'_i$ will be padded with zeros. Therefore, $X'_i$ is further simplified as

$$X'_i(\kappa_1) = (S_{sub} - l) \cdot X_i(\kappa_1) e^{j2\pi f_d^* iTS_{sym}} e^{-\frac{j2\pi N_\tau \kappa_1}{N}}, \quad (13)$$

where $\kappa_1 = 0,1,...,N-1$. Thus, the tunable maximum time delay without ISI is $\min(T_{cp} + lT_s/S_{sub}, T_s)$. Note that there is a trade-off between signal energy loss and maximum time delay with different choices of $l$. Larger $l$ can support larger ISI-free time delay estimation but yields lower signal energy for sensing algorithms analysis. Fig. 8 illustrates an instance of $l = S_{sub} - 1$, and $S_{sub} = 4$.

## V. Ambiguity Performance of 2D FFT

For 2D FFT, the conclusion is similar to delay-and-sum, with the exception that side peaks appear at $(0, \pm 1/T)$. Also, 2D FFT does not require time sequence design since the post-FFT sensing algorithm removes received frequency-domain sequence scrambling without correlation. In addition to independence from scrambling sequence properties, 2D FFT can also adopt staggering comb patterns to eliminate side peaks. Based on Eq. (13), the 2D FFT algorithm performs a second FFT operation to a sequence of $M$ descrambled post-FFT OFDM symbols $\{X_{Ri}\}_{i=0}^{M-1}$, where the $\kappa_1$-th element of $X_{Ri}$ is:

$$X_{Ri}(\kappa_1) = \begin{cases} \dfrac{X'_i(\kappa_1)}{(S_{sub}-l)X_i(\kappa_1)} = e^{j2\pi f_d^* i T S_{sym}} e^{-\frac{j2\pi N_\tau \kappa_1}{N}}, & X'_i(\kappa_1) \neq 0 \\ 0, & \text{otherwise} \end{cases} \quad (14)$$

for $\kappa_1 = 0,1,...N-1$. The 2D FFT result is written as:

$$P(g,q) = \left| \sum_{\kappa_1=0}^{N-1} \left( \sum_{i=0}^{M-1} X_{Ri}(\kappa_1) e^{-\frac{j2\pi q i S_{sym}}{M}} \right) e^{\frac{j2\pi g \kappa_1}{N}} \right|^2$$

$$= \left| \sum_{k=0}^{\frac{N}{S_{sub}}-1} e^{\frac{j2\pi(g-N_\tau)kS_{sub}}{N}} \sum_{i=0}^{M-1} e^{j2\pi f_d^* i T S_{sym}} e^{-\frac{j2\pi q i S_{sym}}{M}} e^{\frac{j2\pi(g-N_\tau)F_i}{N}} \right|^2 \quad (15)$$

For the first summation, the maximum value is obtained at $e^{-\frac{j2\pi N_\tau k S_{sub}}{N}} e^{\frac{j2\pi g k S_{sub}}{N}} = 1$. That is, for

$$\frac{g S_{sub}}{N} - \frac{N_\tau S_{sub}}{N} = z_1 \in \mathbb{Z}, \quad (16)$$

at the 2-D FFT result there will be side peaks at the discrete-valued time delay domain, namely $gT_s/N = \tau$-axis. The side peaks appear at $\tau = \frac{gT_s}{N} = \frac{N_\tau T_s}{N} + \frac{z_1 T_s}{S_{sub}}$. Substituting $g = \frac{Nz_1}{S_{sub}} + N_\tau$, terms of the second summation of Eq. (15) become

$$\sum_{i=0}^{M-1} e^{j2\pi f_d^* i T S_{sym}} e^{-\frac{j2\pi q i S_{sym}}{M}} e^{-\frac{j2\pi N_\tau F_i}{N}} e^{\frac{j2\pi \left(\frac{Nz_1}{S_{sub}}+N_\tau\right)F_i}{N}}$$

$$= \sum_{i=0}^{M-1} e^{j2\pi f_d^* i T S_{sym}} e^{-\frac{j2\pi q i S_{sym}}{M}} e^{\frac{j2\pi z_1 F_i}{S_{sub}}}. \quad (17)$$

Eq. (17) is the Fourier transform of $e^{\frac{j2\pi z_1 F_i}{S_{sub}}}$ evaluated at $f_d^* T S_{sym} - \frac{q S_{sym}}{M}$, and thus different choices of $F_i$ yield different ambiguity peaks.

Taking the example of $F_i$ being constant over $i$, terms of the summation in Eq. (17) will reach the maximum when

$$\frac{q}{MT} = f_d^* - \frac{z_2}{S_{sym}T}, \quad (18)$$

that is, $e^{j2\pi f_d^* i T S_{sym}} e^{-\frac{j2\pi q i S_{sym}}{M}} = 1$. The delay and Doppler ambiguities in $\frac{q}{MT} = f_d$-axis (Doppler domain) and $\tau$-axis (delay domain) are $\left(f_d^* - \frac{z_2}{S_{sym}T}\right)$ and $\left(\frac{z_1 T_s}{S_{sub}} + \tau^*\right)$, respectively, where $z_1 \neq 0$, $z_2 \neq 0$ and $\tau^* = \frac{N_\tau T_s}{N}$. Without loss of generality, we assume the true time delay and Doppler frequency pair as $(\tau^*, f_d^*) = (0,0)$ in the analysis.

Fig. 9(a) shows the ambiguity performance of $S_{sub} = 4$, $S_{sym} = 1$, and $F_i \equiv 0$ (Scheme A). Similar to delay and sum, the maximum unambiguous delay is restricted to $T_s/S_{sub}$. The maximum unambiguous Doppler is smaller than delay-and-sum due to side peaks at $(0, \pm 1/T)$.

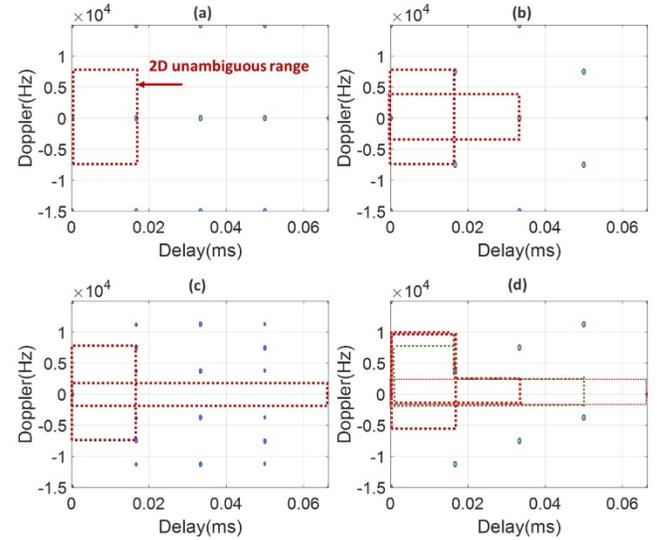

Figure 9. AF of 2D FFT with different RS patterns

Moreover, when $F_i$ is not linearly changing with $i$ (e.g., 5G PRS), the Fourier transform generates several side peaks and results can be calculated numerically. Fig. 9(b) and (c) show 2D maximum unambiguity in the case of $S_{sub} = 4$ and $S_{sym} = 1$, for Scheme B and C, respectively.

When the proposed staggering offset is adopted, i.e., Scheme D, the Fourier transform yields delta functions. Substituting $F_i = p \cdot i$ into terms of the summation, it will reach the maximum when $e^{j2\pi f_d^* i T S_{sym}} e^{-\frac{j2\pi q i S_{sym}}{M}} e^{\frac{j2\pi z_1 pi}{S_{sub}}} = 1$, that is, $f_d^* T S_{sym} - \frac{q S_{sym}}{M} + \frac{z_1 p}{S_{sub}} = z_2 \in \mathbb{Z}$. Here $p$ is relatively prime to $S_{sub}$. This can be further written as

$$\frac{q}{MT} = f_d^* - \frac{z_2}{S_{sym}T} + \frac{z_1 p}{S_{sub}S_{sym}T}. \quad (19)$$

Thus, delay and Doppler ambiguities in $\tau$-axis and $f_d$-axis are $MT\left(f_d^* - \frac{z_2}{S_{sym}T} + \frac{z_1 p}{S_{sub}S_{sym}T}\right)$ and $\left(\frac{z_1 T_s}{S_{sub}} + \tau^*\right)$, respectively, where $z_2 \neq 0$, $\tau^* = \frac{N_\tau T_s}{N}$. Fig. 9(d) presents its 2D maximum unambiguity in the case of $S_{sub} = 4$, $S_{sym} = 1$, and $(\tau^*, f_d^*) = (0,0)$ with the proposed Scheme D. Similarly, Scheme D is also suggested for 2D FFT due to flexibility in tuning 2D maximum unambiguity.

## VI. CONCLUSION

In summary, we investigated general uniformly-spaced, comb RS patterns with various staggering offsets for bi-static integrated communication and sensing systems. Moreover, we proposed an extended guard interval design for comb RS structure when using post-FFT sensing algorithms to support long-range sensing with ISI. We have characterized ambiguity performances of these RS patterns using delay-and-sum and 2D FFT algorithms. A staggering scheme with linear slope relatively prime to the comb size offers more flexible choices of 2D unambiguity under the used algorithms. The results can be applied to bistatic sensing to extend the sensing distance and flexibly adjust distance-velocity unambiguity with RS patterns stemming from existing communication system.